# Boron nitride substrates for high quality graphene electronics

C.R. Dean[1,2], A.F. Young[3], I. Meric[1], C. Lee[2], L. Wang[2], S. Sorgenfrei[1],

K. Watanabe[4], T. Taniguchi[4], P. Kim[3], K.L. Shepard[1], J. Hone[2]

[1]*Department of Electrical Engineering,*

*Columbia University, New York, NY, 10027, USA*

[2]*Department of Mechanical Engineering,*

*Columbia University, New York, NY, 10027, USA*

[3]*Department of Physics, Columbia University, New York, NY, 10027, USA and*

[4]*Advanced Materials Laboratory, National Institute for Materials Science,*

*1-1 Namiki, Tsukuba, 305-0044, Japan*

Graphene devices on standard SiO$_2$ substrates are highly disordered, exhibiting characteristics far inferior to the expected intrinsic properties of graphene[1–12]. While suspending graphene above the substrate yields substantial improvement in device quality[13,14], this geometry imposes severe limitations on device architecture and functionality. Realization of suspended-like sample quality in a substrate supported geometry is essential to the future progress of graphene technology. In this Letter, we report the fabrication and characterization of high quality exfoliated mono- and bilayer graphene (MLG and BLG) devices on single crystal hexagonal boron nitride (h-BN) substrates, by a mechanical transfer process. Variable-temperature magnetotransport measurements demonstrate that graphene devices on h-BN exhibit enhanced mobility, reduced carrier inhomogeneity, and reduced intrinsic doping in comparison with SiO$_2$-supported devices. The ability to assemble crystalline layered materials in a controlled way sets the stage for new advancements in graphene electronics and enables realization of more complex graphene heterostructres.

The quality of substrate-supported graphene devices has not improved since the first observation of the anomalous quantum Hall effect in graphene and its bilayer[1,2]. On SiO$_2$, the carrier mobility is limited by scattering from charged surface states and impurities[3–6,8], substrate surface roughness[9–11] and SiO$_2$ surface optical phonons[7,8]. Moreover, near the Dirac point substrate-induced disorder breaks up the 2D electron gas (2DES) into an inhomogeneous network of electron and hole puddles[5,6,12], while charged impurities trapped in the substrate or at the graphene-substrate interface cause doping of the 2DES away from charge neutrality. So far, efforts to engineer alternatives to SiO$_2$ have typically involved other oxides, where similar surface effects continue to be problematic[17–19].

Hexagonal boron nitride (h-BN) promises to be an ideal substrate dielectric for improved graphene-based devices. h-BN is an insulating isomorph of graphite with boron and nitrogen atoms occupying the inequivalent $A$ and $B$ sublattices in the Bernal structure. The different onsite energies of the B and N atoms lead to a large (5.97 eV) band gap[20] and a small (1.7%) lattice mismatch with graphite[21]. Owing to the strong, in-plane, ionic bonding of the planar hexagonal lattice structure, h-BN is relatively inert and expected to be free of dangling bonds or surface charge traps. Furthermore, the atomically planar surface should suppress rippling in graphene, which has been shown to mechanically conform to both corrugated and flat substrates[9,22]. The dielectric properties of h-BN ($\epsilon \sim 3-4$ and $V_{\text{Breakdown}} \sim 0.7$ V/nm) compare favorably with SiO$_2$, allowing the use of h-BN as a an alternative gate dielectric with no loss of functionality[15]. Moreover, the surface optical phonon modes of h-BN have energies two times larger than similar modes in SiO$_2$, suggesting the possibility of improved high-temperature and high-electric field performance of h-BN based graphene devices over those using typical oxide/graphene stacks[23].

To fabricate graphene-on-BN, we employ a mechanical transfer process, illustrated in Fig. 1 (see Methods). The h-BN flakes used in this study are exfoliated from ultra-pure, hexagonal-BN single crystals, grown by the method described in Ref. 24. The optical contrast on 285 nm SiO$_2$/Si substrates is sufficient to easily identify h-BN flakes with thicknesses down to a single monolayer (see Fig. 1b as well as Ref. 16). Fig. 2 shows AFM images of MLG transferred onto $\sim$14 nm thick h-BN (see also supplementary information (SI)). The transferred graphene is free of wrinkles or distortions, consistent with previous reports of similar polymethyl-methacrylate- (PMMA) based transfer techniques[25]. A histogram of the roughness of graphene on h-BN (Fig. 2b) shows it to be indistinguishable from bare h-BN and approximately three times less rough than SiO$_2$. We conclude that the graphene membrane conforms to the atomically flat h-BN, consistent with previous reports on both rippled[9] and flat[22] surfaces.

Electronic transport measurements of MLG transferred onto h-BN indicate that the resulting two-dimensional electronic systems are of high quality. Fig. 3a shows the resistance of a typical MLG sample on h-BN as a function of applied back gate voltage, $V_g$. The resistivity peak, corresponding to the overall charge neutrality point (CNP), is extremely narrow and occurs at nearly zero gate voltage. The conductivity (dotted line inset in Fig. 3a) is strongly sublinear in carrier density, indicating a crossover from scattering dominated by charge impurities at low density to short-range impurity scattering at large carrier density[4–6,11,26]. The data are well fit by a self-consistent Boltzmann equation for diffusive transport that includes both long and short range scattering (solid line in figure)[5,6], $\sigma^{-1} = (ne\mu_C + \sigma_o)^{-1} + \rho_s$, where $\mu_C$ is the density-independent mobility due to charged-impurity Coulomb (long-range) scattering, $\rho_S$ is the contribution to resistivity from short-range scattering, and $\sigma_o$ is the residual conductivity at the CNP. We obtain $\mu_C \sim 60,000$ cm$^2$/Vs, three times larger than on SiO$_2$ using a similar analysis[26], and $\rho_S \sim 71$ $\Omega$, which is similar to values obtained on SiO$_2$. This indicates a threefold decrease in the scattering rate due to charge-impurities in this sample, but a similar degree of short range scattering, in comparison to the best $SiO_2$ samples. This suggests that the sublinear shape does not result from increased short range scattering on BN substrates, but rather a substantially reduced charge impurity contribution, which reveals the effects of short range scattering at comparatively lower densities. Similar behavior was observed in more than 10 MLG samples and, importantly, we always measure a higher mobility for BN-supported graphene as compared to portions of the same flake on the nearby $SiO_2$ surface (see SI). For the MLG device shown here, the Hall mobility is ~25,000 at high density, where short range scattering appears to dominate. While the origin of short-range scattering remains controversial, the similar values of $\rho_S$ between $SiO_2$ and h-BN supported-graphene samples suggests that scattering off ripples and out-of-plane vibrations[10,11] may not be a significant contribution in our samples since these are likely to be suppressed on atomically flat h-BN.

The width of the resistivity peak at the CNP gives an estimate of the charge-carrier inhomogeneity resulting from electron-hole puddle formation at low density[28]. In Fig. 3a the full width at half maximum (FWHM) of $\rho(V_g)$ is ~1 V, giving an upper bound for disorder-induced carrier density fluctuation of $\delta n < 7 \times 10^{10}$ cm$^{-2}$, a factor of ~3 improvement over $SiO_2$-supported samples[12]. An alternate estimate of this inhomogeneity is obtained from the temperature dependence of the minimum conductivity. In Fig. 3c, $\sigma_{min}$ increases by a factor of two between 4 K and 200 K. Such a strong temperature dependence has previously only been observed in suspended samples, with substrate-supported samples typically exhibiting $< 30\%$ variation in the same range[13]. $\sigma_{min}$ is expected to vary with temperature only for $k_B T > E_{\text{puddle}}$ where for MLG[13] $E_{\text{puddle}} \approx \hbar v_f \sqrt{\pi \delta n}$. Here $\sigma_{min}$ saturates to $\sim 6e^2/h$ for $T \lesssim 15$ K giving an upper bound of $\delta n \sim 10^9$ cm$^{-2}$. The $\delta_n$ estimated by these two measures is consistent with similar analysis performed on suspended devices[13,29].

It has been proposed that a bandgap would be induced in graphene aligned to an h-BN substrate[21]. In our experiment the graphene has a random crystallographic orientation to the substrate, and thus we do not expect the necessary symmetry breaking to occur. Indeed, the temperature dependence of $\sigma_{min}$ observed here does not follow a simply activated behavior, suggesting no appreciable gap opening in this randomly stacked graphene on h-BN.

Transport measurements from BLG transferred to h-BN are shown in Fig. 3b. The corresponding conductivity is linear in gate voltage up to large densities, as expected for BLG in the presence of long and short range scalar potential disorder[30]. The (density-independent) electron and hole Hall mobilities are ∼60,000 cm$^{-2}$/Vs and ∼80,000 cm$^{-2}$/Vs, respectively, at $T =2$ K, with a value of 40,000 cm$^{-2}$/Vs measured at room temperature in air for this same device. The FWHM of the CNP resistivity peak is ∼1.2 V, giving an estimate of the carrier inhomogeneity density $\delta n \sim 9 \times 10^{10}$ cm$^{-2}$. Both the mobility and inhomogeneity are comparable to the best reported suspended BLG devices[29] and almost an order of magnitude better than BLG on SiO$_2$[11]. The temperature dependence of $\sigma_{min}$ (blue circles in Fig. 3c) is much stronger than in MLG, consistent with previous studies[11,26] (We note that the BLG studied here, although undoped immediately after sample fabrication and annealing, was contaminated upon insertion into a Helium flow cryostat; thereafter the CNP was found at $V_g \sim -27$ V. The temperature dependence at the CNP may therefore be due in part to an electric field induced energy gap[31,32]).

The temperature dependence of the resistivity at high density for both MLG and BLG is shown in Fig. 3d. MLG resistance increases linearly with temperature (solid line in Fig. 3d) due to longitudinal acoustic (LA) phonon scattering, $\rho_{LA}(T) = \left(\frac{h}{e^2}\right) \frac{\pi^2 D_A^2 k_B T}{2h^2 \rho_s v_s^2 v_f^2}$, where $\rho_s = 7.6 \times 10^{-7}$ kg/m$^{-2}$ is the graphene mass density, $v_f = 1 \times 10^6$ m/s is the Fermi velocity, $v_s = 2 \times 10^4$ m/s is the LA phonon velocity and $D_A$ is the acoustic deformation potential[8,13]. Linear fits to the electron (hole) branches give $D_A \sim 18$ eV ($D_A \sim 21$ eV). In contrast, BLG exhibits a very weak temperature dependence, with a slightly negative overall trend (dashed line in Fig. 3d). Both of these findings agree with previous measurements[8,11,13,26]. We note that no indication of activated remote surface phonon scattering is seen in MLG (BLG) up to 200 K (240 K). However, further studies in a variable temperature UHV environment[8] are need to explore the high temperature behavior in graphene-on-BN more fully.

The replacement of the SiO$_2$ substrate with h-BN appears to result in a marked change in the chemical properties of graphene devices. Fig. 3e shows the room-temperature conductivity of a typical MLG layer before and after annealing in a H$_2$/Ar flow at 340°C for 3.5 hrs (see methods). Annealing substantially enhances the carrier mobility while leaving the position of the CNP virtually unchanged. The low mobility immediately post-transfer may be due to neutral transfer residues and/or local strains that are relaxed upon heating. The lack of doping after heating in H$_2$/Ar is in stark contrast to SiO$_2$-supported devices, where heat treatment typically results in heavy doping of the graphene, often more than $5 \times 10^{12}$cm$^{-2}$, after re-exposure to air. We speculate that the reduced chemical reactivity of graphene on h-BN is due to a combined effect of the chemically inert and gas-impermeable h-BN surface together with reduced roughness in the graphene film.

Magnetotransport measurements provide further confirmation of the high material quality achieved in these samples. Fig. 4a shows the magnetoconductivity $\sigma_{xx}$ and Hall conductivity $\sigma_{xy}$ as a function of density at B=14 T for MLG, derived from simultaneous measurement of magnetoresistance $R_{xx}$ and Hall resistance $R_{xy}$ in the Hall bar geometry shown in Fig. 2. Complete lifting of the four-fold degeneracy[27] of the zero energy Landau level (LL) is observed, with the additional quantum hall states at $\nu = 0, +1, \pm 2$ exhibiting quantized Hall conductance $\sigma_{xy} = \nu e^2/h$ together with vanishing $\sigma_{xx}$. The dashed line in Fig. 4a indicates that signatures of the $\nu = \pm 1$ quantum hall effect (QHE) are visible at fields as low as $B = 8.5$ T, more than a factor of two smaller than reported for MLG on $SiO_2$[27].

A complete sequence of broken symmetry LLs are visible in BLG $B = 14$ T (Fig. 4b). In our device, the substrate supported geometry allows us to probe much higher density than possible in suspended devices of similar quality[29]. Quantized Hall resistance is observed at $R_{xy} = \frac{1}{\nu} h/e^2$ concomitant with minima in $R_{xx}$ for all integer filling factors from $\nu = 1$ to at least $\nu = 16$. Density sweeps at lower fields (see SI) show that the lifting of the expected four-fold degeneracy in BLG[29] is observable up to at least the fifth LL at fields as low as 5 T. Complete quantization of the four-fold degenerate LLs and Shubnikov-de Haas oscillations seen down to 2 and 0.4 T respectively. (see inset in Fig. 4b and also SI).

In the lowest LL the $\nu = 2$ quantum Hall state has a larger gap compared to the states at $\nu = 1$ and 3, as judged by the depth of the $R_{xx}$ minimum. Interestingly, in the second LL, the situation is reversed, with $\nu = 6$ weaker than $\nu = 5, 7$. As the LL index is increased, the trend in the gaps evolves back towards that observed in the lowest LL. A full understanding of symmetry breaking with increasing LL index is complicated by the fact that the applied gate voltage and residual extrinsic doping can both simultaneously break the layer degeneracy in BLG and modify the exchange energy. Analysis of this trend is, therefore, left to a future study in dual gated devices where the transverse electric field can be tuned independently. Preservation of high mobility in dual-gated device may be achieved by fabricating h-BN–graphene–h-BN stacks using a two-transfer technique[15].

## I. METHODS

Graphene-on-BN devices were fabricated according to the procedure illustrated in (Fig. 1d): (i) Fabrication begins with the mechanical exfoliation of h-BN single crystals onto silicon wafers coated in 285 nm thermal oxide. Graphene is exfoliated separately onto a polymer stack consisting of a water soluble layer (Mitsubishi Rayon aquaSAVE) and PMMA, and the substrate is floated on the surface of a DI water bath; (ii) Once the water-soluble polymer dissolves, the Si substrate sinks to the bottom of the bath leaving the extremely hydrophobic PMMA floating on top, (iii) The PMMA membrane is adhered to a glass transfer slide, which is clamped onto the arm of a micromanipulator mounted on an optical microscope. Using the microscope the graphene flake is precisely aligned to the target BN and the two are brought into contact. During transfer, the target substrate is heated to 110 °C in an effort to drive off any water adsorbed on the surface of the graphene or h-BN flakes as well as to promote good adhesion of the PMMA to the target substrate; (iv) Once transferred, the PMMA is dissolved in acetone. Electrical leads are deposited using standard electron beam lithography, after which all our samples are annealed in flowing $H_2$/Ar gas at 340 °C for 3.5 hours to remove resist residues. The devices presented in the main text did not undergo any further treatment (i.e. in-situ vacuum annealing etc.) after removal from the $H_2$/Ar flow.

AFM images were acquired in air using silicon cantilevers operated in tapping mode. Surface roughness is reported as the standard deviation of the surface height distribution (determined by a fitted Gaussian), measured on a 0.3 $\mu m^2$ area. Transport measurements were acquired in a four-terminal geometry using standard lock-in techniques at ∼17 Hz. Samples were cooled in a variable temperature (∼2–300 K) liquid $^4$He flow cryostat with the sample in vapor.

## II. ACKNOWLEDGMENTS

We thank D. Sukhdeo and N. Baklitskaya for help with the device fabrication. This work is supported by DARPA CERA, AFOSR MURI, ONR MURI, FCRP through C2S2 and FENA, NSEC (No. CHE-0117752) and NYSTAR

## III. AUTHOR CONTRIBUTIONS

CRD and AFY performed the experiments including sample fabrication, measurement, characterization, and development of the transfer technique. IM contributed to sample fabrication and measurement. CL, and WL, contributed to sample fabrication. SS contributed to development of the transfer technique. KW and TT synthesized the h-BN samples. PK, KLS, and JH advised on experiments.

**FIGURES**

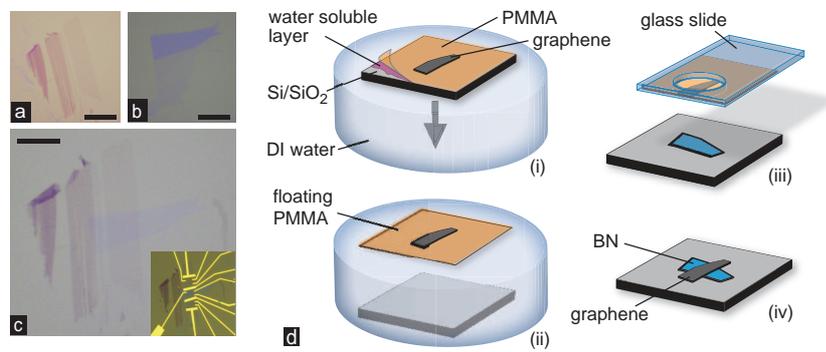

FIG. 1. Optical images of graphene and h-BN before (a and b, respectively) and after (c) transfer. Scale bar in each is 10 $\mu$m. Inset shows electrical contacts. (d) Schematic illustration of the transfer process to fabricate graphene-on-BN devices (see text for details).

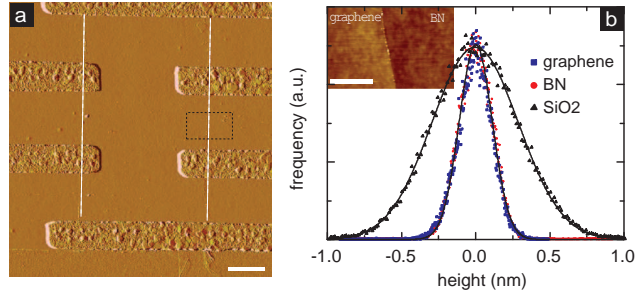

FIG. 2. (a)AFM image of monolayer graphene on BN with electrical leads. White dashed lines indicate the edge of the graphene flake. Scale bar is 2 $\mu$m. (b) Histogram of the height distribution (surface roughness) measured by AFM for SiO$_2$ (black triangles), h-BN (red circles) and graphene-on-BN (blue squares). Solid lines are Gaussian fits to the distribution. Inset: high resolution AFM image showing comparison of graphene and BN surfaces, corresponding to the dashed square in (a). Scale bar is 0.5 $\mu$m

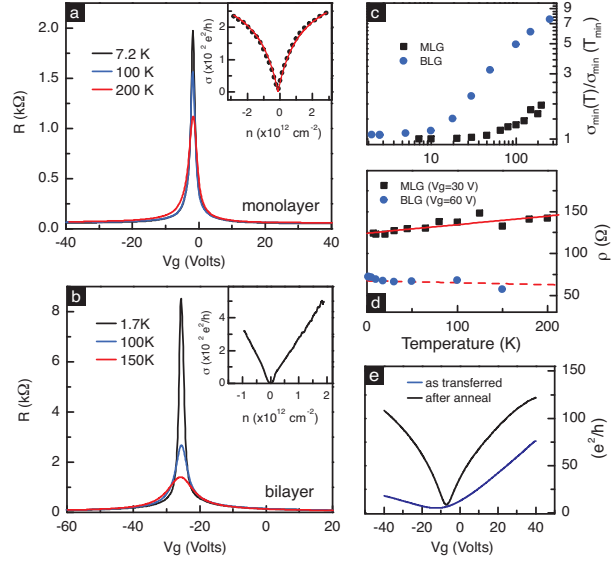

FIG. 3. Resistance versus applied gate voltage for (a) MLG and (b) BLG on h-BN. Inset in each panel shows the corresponding conductivity. For both devices, the temperature dependence of the conductivity minimum and high density resistivity are shown in (c) and (d), respectively. Solid and dashed lines in (d) are linear fits to the data. (e) Conductivity of a different MLG sample comparing the room-temperature transport characteristics measured as–transferred–to–h-BN (blue curve) and after annealing in $H_2Ar$ (black curve).

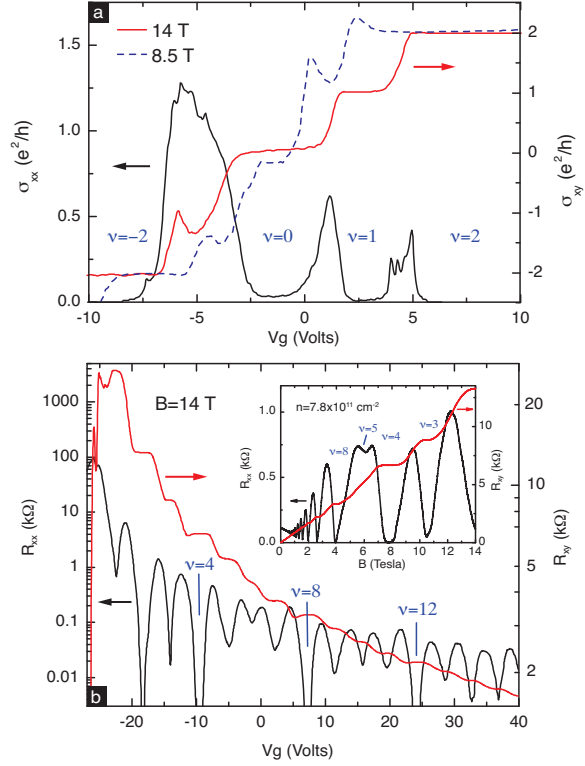

FIG. 4. (a)Longitudinal and Hall conductivity versus gate voltage at $B = 14$ T (solid line) and 8.5 T (dashed line) for MLG. (b) Longitudinal and Hall resistance versus gate voltage at $B = 14$ T for BLG. Inset shows a magnetic field sweep at fixed density. SdH oscillations begin at ∼0.4 T with LL symmetry breaking appearing at fields less than 6 T. $T \sim 2$ K in both panels.

# Boron nitride substrates for high quality graphene electronics: Supplementary Information

C.R. Dean[1,2], A.F. Young[3], I. Meric[1], C. Lee[2], W. Lei[2], S. Sorgenfrei[1],
K. Watanabe[4], T. Taniguchi[4], P. Kim[3], K.L. Shepard[1], J. Hone[2]

[1]*Department of Electrical Engineering,*

*Columbia University, New York, NY, 10027, USA*

[2]*Department of Mechanical Engineering,*

*Columbia University, New York, NY, 10027, USA*

[3]*Department of Physics, Columbia University, New York, NY, 10027, USA and*

[4]*Advanced Materials Laboratory, National Institute for Materials Science,*

*1-1 Namiki, Tsukuba, 305-0044, Japan*

# I. AFM CHARACTERIZATION OF H-BN

Before transferring graphene, the surface of every target h-BN flake is first characterized by atomic force microscopy to ensure it is free of contaminants or step edges, and also to measure its thickness. Fig. SS1a-b shows an example optical and AFM image of a clean h-BN surface after transfer onto a $SiO_2$ substrate. While the texture of the $SiO_2$ surface is visibly apparent, the h-BN surface looks completely devoid of any features on this scale.

Fig. SS1c shows a histogram of the measured surface roughness for h-BN flakes of varying thicknesses. Measurements from a typical $SiO_2$ substrate, and from a calibration HOPG wafer are also shown, for comparison. All data was acquired on a 300 $nm^2$ scan window. The $SiO_2$ surface roughness , given by the standard deviation of a fitted Gaussian, is measured

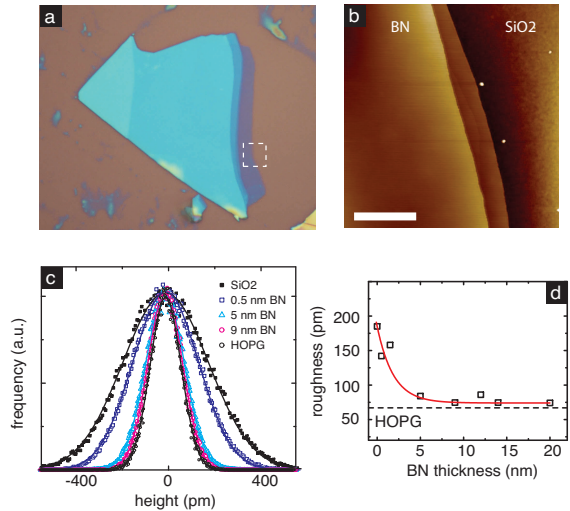

FIG. S1: (a)Optical image of a representative h-BN flake exfoliated onto a $Si/SiO_2$ substrate. (b) AFM image of the region indicated in (a) by a dashed box. scale-bar is 0.5 $\mu$m. The h-BN surface seen here measures $\sim$ 8 nm in height relative to the $SiO_2$ backgraound. At this scale it is apparent the h-BN surface is much smother than the underlying $SiO_2$ substrate. (c) Height histogram of the h-BN surface measured for several different sample-thicknesses. A typical measurement from a $SiO_2$ surface (solid black squares) and a HOPG wafer (open black circles) are shown for comparison. (d) h-BN surface roughness versus sample thickness measured from several different samples. Solid line is a guide-to-the-eye. Dashed line indicates resolution of our system, obtained by measuring the surface of HOPG under the same conditions.

to be ∼ 185 pm, consistent with values reported elsewhere[1]. The HOPG surface roughness is ∼ 70 pm, which, since the HOPG wafer is atomically flat over large areas, is taken to be the resolution limit of our measurement. As seen in Fig. SS1d, the h-BN surface roughness approaches the measured HOPG roughness for flakes thicker than approximately 5 nm.

## II. COMPARISON BETWEEN DIFFERENT SAMPLES

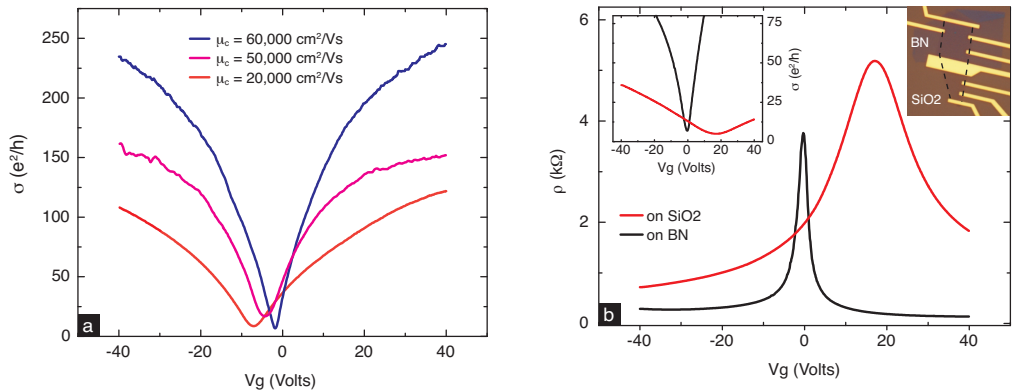

FIG. S2: (a)Representative conductivity curves measured for three different MLG samples transferred to h-BN. Legend indicates the corresponding mobility extracted by fitting to the Boltzmann model from the main paper. (b) Resistivity measured on a single flake spanning both BN and $SiO_2$ substrate regions. Inset left shows corresponding conductivities. Inset right shows optical image of the sample where the dashed line outlines the graphene. $T \sim 4$ K in both (a) and (b).

Fig. SS2a shows conductivity curves measured from three representative MLG layers transfered to h-BN. The mobilities indicated in the figure are extracted from fits using the same equation as in the main text. Similar to what has been reported on $SiO_2$, there appears to be a correlation between sample quality and the charge neutrality position as well as the the width of the conductivity minimum[2,3]. Specifically, high quality samples coincide with a sharply defined conductivity minimum occurring near zero backgate, whereas, poorer quality samples exhibit broader minima further away from zero backgate voltage. This is consistent with the mobility enhancement observed in graphene–on–h-BN resulting from a reduction of charged impurities, relative to graphene on $SiO_2$. Further evidence of this is shown in Fig. SS2b, where portions of the same graphene flake are measured both on h-BN and on $SiO_2$.

While on h-BN the graphene exhibits a very narrow resistivity peak, occurring nearly at zero gate voltage, on SiO$_2$ the same flake is significantly doped ($V_{CNP} \sim 25$ Volts), and shows a broad peak. From the corresponding conductivity curves (shown inset in the figure) we measure a mobility for the h-BN and SiO$_2$ supported regions of *the same graphene flake* to be $\sim 20,000$ cm$^2$/Vs and $\sim 2,000$ cm$^2$/Vs, respectively. While variation in sample quality, within the same graphene flake, is observed[2] on samples supported only by SiO$_2$, we always observe a higher mobility on h-BN relative to SiO$_2$, when measuring a portion of the same flake on both surfaces.

## III.  MAGNETOTRANSPORT IN BILYAER GRAPHENE

Fig. SS3a shows an enlargement of the magneto-transport measured from BLG on h-BN presented in Fig. 4 of the main text. Landau levels are labeled between 5 and 14 Tesla, indicating that appearance of the four-fold symmetry breaking is visible down to approximately 5 Tesla. Complete quantization of the four-fold degenerate LL's, evidence by both quantization in $R_{xy}$ and a zero value minimum in $R_{xx}$, is observed down to approximately 2 Tesla. The inset of Fig. SS3a shows the low field Shubnikov de Haas oscillations, which are visible down to as low as 0.4 Tesla.

Magnetoresistance measured at fixed field, but varying backgate voltage, are shown for several different fields in Fig. SS3b. Minima in between the otherwise four-fold degenerate LL's, for LL index greater than $\nu = 4$, begin to emerge at $\sim$5 Tesla, becoming fully quantized for all integer fillings up to at least $\nu = 20$ at 14 Tesla. In the lowest energy LL, where the n=0 and n=1 levels are doubly degenerate, the $\nu = 2$ quantum Hall state shows a deep broad minimum at fields well below 5 Tesla.

---

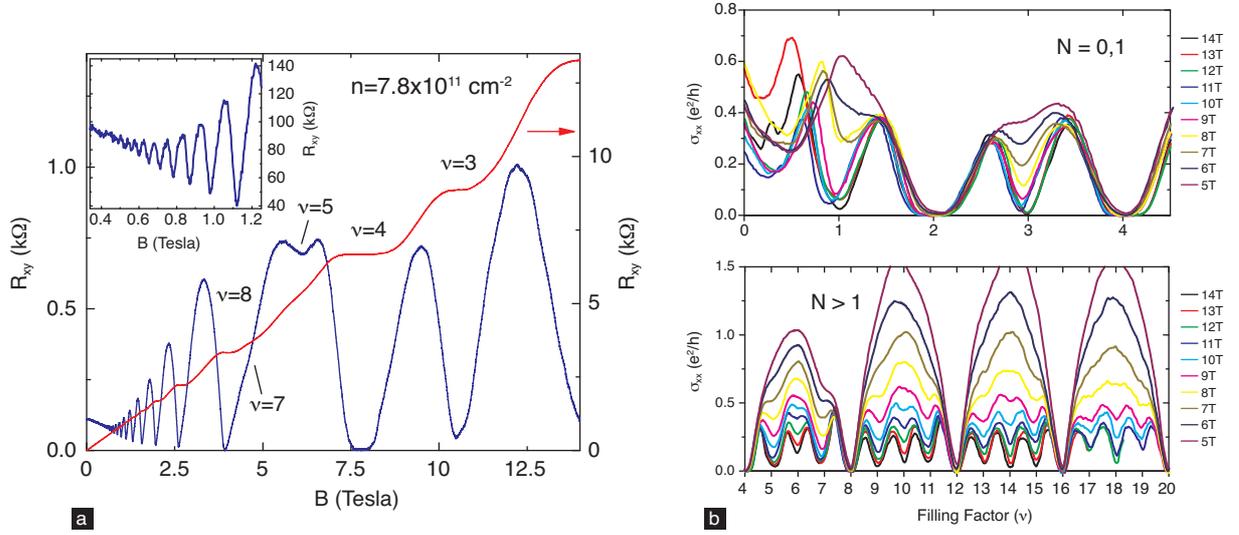

FIG. S3: (a) Magnetoresistance (blue curve) and Hall resistance (red curve) versus B field of the BLG sample on h-BN. $T \sim 4$ K and $n = 7.8 \times 10^{11}$ cm$^{-2}$. Landau Levels between 5 and 14 Tesla are labeled. Inset shows low field SdH oscillations, measured under the same conditions. (b) Magnetoresistance versus gate voltage of the same sample. Upper panel shows symmetry breaking in the lowest energy Landau Level (i.e. $|\nu| < 4$). Lower panel shows symmetry breaking of the higher order Landau levels. The data is plotted versus filling factor for easier comparisons between different magnetic fields.